\documentclass{article}
\usepackage[preprint]{spconfa4}
\usepackage{amsmath,amssymb,mathtools,graphicx}
\usepackage{flushend}

\usepackage[backend=biber, bibencoding=utf8,
            style=ieee,
            maxbibnames=99,
            doi=false,
            url=false,
            isbn=false]{biblatex}
\AtEveryBibitem{%
  \ifentrytype{inproceedings}{%
    \clearfield{pages}%
    \clearlist{publisher}%
    \clearlist{organization}}{} }

\usepackage{booktabs}
\usepackage{multirow}
\usepackage{siunitx}
\usepackage{microtype}
\usepackage[hidelinks]{hyperref}
\usepackage[capitalise]{cleveref}
\crefname{equation}{}{}

\newcommand{\setA}{\mathcal{A}}
\newcommand{\setC}{\mathcal{C}}

\newcommand{\pA}[0]{P_\setA}
\newcommand{\pC}[0]{P_\setC}

\newcommand{\rA}[0]{R_\setA}
\newcommand{\rC}[0]{R_\setC}

\newcommand{\fA}[0]{f_\setA}
\newcommand{\fC}[0]{f_\setC}

\newcommand{\mat}[1]{\mathbf{#1}}

\newcommand{\A}[0]{\mat{A}}
\newcommand{\X}[0]{\mat{X}}
\newcommand{\Y}[0]{\mat{Y}}
\newcommand{\U}[0]{\mat{U}}
\newcommand{\Z}[0]{\mat{Z}}
\newcommand{\W}[0]{\mat{W}}
\newcommand{\PH}[0]{\mat{\Phi}}

\DeclareMathOperator{\STFT}{STFT}
\DeclareMathOperator{\iSTFT}{iSTFT}

\newcommand{\eqcomma}{\,,} 
\newcommand{\eqperiod}{\,.} 

\DeclarePairedDelimiter{\abs}{\lvert}{\rvert}

\title{Beyond Griffin-Lim: Improved Iterative Phase Retrieval for Speech}
\name{Tal Peer$^{1}$, Simon Welker$^{1,2}$, Timo Gerkmann$^{1}$\thanks{\scriptsize This work was funded by the Deutsche Forschungsgemeinschaft (DFG, German Research Foundation) --- project number 247465126, as well as by DASHH (Data Science in Hamburg - HELMHOLTZ Graduate School for the Structure of Matter) --- Grant-No. HIDSS-0002.}}
\address{$^{1}$ Signal Processing (SP), Universität Hamburg, Germany \\
      $^{2}$ Center for Free-Electron Laser Science, DESY, Hamburg, Germany}
\begin{document}
\ninept
\maketitle
\begin{abstract}
Phase retrieval is a problem encountered not only in speech and audio processing, but in many other fields such as optics. Iterative algorithms based on non-convex set projections are effective and frequently used for retrieving the phase when only STFT magnitudes are available. While the basic Griffin-Lim algorithm and its variants have been the prevalent method for decades, more recent advances, e.g. in optics, raise the question: Can we do better than Griffin-Lim for speech signals, using the same principle of iterative projection?

In this paper we compare the classical algorithms in the speech domain with two modern methods from optics with respect to reconstruction quality and convergence rate. Based on this study, we propose to combine Griffin-Lim with the Difference Map algorithm in a hybrid approach which shows superior results, in terms of both convergence and quality of the final reconstruction.
\end{abstract}
\section{Introduction}
\vspace{-2px}
\label{sec:intro}
While recorded audio (including speech) is naturally represented as a time-domain signal, many speech processing systems do not operate directly in the time-domain but rather transform the signal first into a time-frequency representation, typically with the short-time Fourier transform (STFT). In contrast to the real-valued time-domain signal, the STFT representation is complex-valued and its magnitude and phase components are often treated separately. While phase-aware approaches are becoming increasingly common in the fields of speech enhancement and source separation, many algorithms only modify the STFT magnitude and retain the input phase \cite{gerkmannPhaseProcessingSingleChannel2015a}. In other tasks, such as speech synthesis \cite{takakiDirectModelingFrequency2017} or time-scale modification \cite{griffinSignalEstimationModified1984}, the phase is sometimes completely missing. In both cases, application of the inverse STFT (iSTFT) requires both magnitude and phase components. Thus, one needs to \emph{retrieve} the phase component, which must be consistent to some extent with the magnitude component.

The phase retrieval problem is not unique to speech or audio. It appears in many fields which use techniques involving a Fourier-like transformation of data \cite{shechtmanPhaseRetrievalApplication2015b}. A prominent example is coherent diffractive imaging, where measured wave intensities (squared amplitudes) correspond to diffraction patterns which can be Fourier-transformed into object images, given the missing phase component. An imaging technique that is particularly similar to STFT processing is ptychography \cite{rodenburgPtychography2019}, where the imaged object is divided into overlapping segments, creating a redundancy that is useful for phase retrieval. This similarity makes the application of ptychographic methods to audio (and vice-versa) straightforward \cite{welkerDeepIterativePhase2022a}.

A typical approach to phase retrieval, in audio as well as in other fields, is the use of iterative projection algorithms, which are based on successive projection of the data onto sets defined by different constraints. While the constraints are chosen based on the problem at hand, the general framework is similar in the different fields, allowing for easy adaptations of algorithms between them. Again, this is especially true in the case of ptychography, where the constraints are very similar to those used in STFT processing.

The de-facto standard for STFT phase retrieval in the context of audio signals is the Griffin-Lim algorithm (GLA) \cite{griffinSignalEstimationModified1984}, which is itself inspired by the earlier Gerchberg-Saxton algorithm \cite{gerchbergPracticalAlgorithmDetermination1972} used in optics. Interestingly, GLA was later rediscovered in ptychography as the Error Reduction algorithm \cite{bauschkePhaseRetrievalError2002,rodenburgPtychography2019}. Several extensions of GLA have been proposed over the years, including an accelerated version \cite{perraudinFastGriffinLimAlgorithm2013}, online variants \cite{beauregard2005efficient,zhuRealTimeIterativeSpectrum2006} an adaptation to the source separation context \cite{gunawanIterativePhaseEstimation2010} and even augmentation by a neural network \cite{masuyamaDeepGriffinLim2021}, but the basic algorithm is still widely used (other approaches not based on iterative projection exist \cite{gerkmannPhaseProcessingSingleChannel2015a} but are outside the scope of this paper). In this work we examine how modern methods from optics perform when applied to speech signals. By increasing the number of algorithms to choose from, we are also able to analyse different combinations of algorithms in a hybrid approach.

During the final stages of work on this paper, parallel work by Kobayashi et al. \cite{kobayashiAcousticApplicationPhase2022} has been published, where the authors also seek to apply modern phase retrieval algorithms to speech signals, and show that in some cases their performance surpasses that of GLA. While the premise of that paper is similar to ours, our contribution goes beyond \cite{kobayashiAcousticApplicationPhase2022} in two aspects. First, we consider an additional important algorithm: the Difference Map~\cite{elserPhaseRetrievalIterated2003}. Second, we propose a novel hybrid algorithm which delivers superior results in terms of speech quality and intelligibility.
\section{Iterative Projection Algorithms}
\vspace{-2px}
\label{sec:algos}
The general idea behind iterative projection algorithms is iterative application of constraints on the signal, converging towards a point in which all constraints are met. The constraints are applied by projecting the signal onto sets defined by the respective constraints. In STFT phase retrieval, we typically consider the magnitude constraint set $\setA$ and a consistency constraint set $\setC$. Since we assume the magnitude is known and only phase is sought, we simply define $\setA$ as the set of all complex spectrograms that have a known magnitude spectrogram $\A$. To project a spectrogram $\X$ onto $\setA$, we simply enforce the known magnitude $\A$:
\begin{equation}
    \label{eq:PA}
    \pA(\X) = \A \frac{\X}{\abs{\X}} \eqcomma
\end{equation}
with multiplication, division and absolute value performed element-wise. This operation is only applied to non-zero STFT bins to avoid division by zero. 

While the magnitude constraint is used in phase retrieval in many fields, the definition of the other constraint is rather problem-dependent and can be chosen based on physical or mathematical characteristics of the signals. In the context of  STFT phase retrieval, we may make use of the inherent redundancy of the STFT representation (due to overlapping frames), and define the constraint in terms of \emph{consistency}, i.e. $\setC$ is defined as the set of all spectrograms that correspond to a time-domain signal \cite{gerkmannPhaseProcessingSingleChannel2015a}. The projections onto this set is achieved by successively transforming the signal into the time-domain and back to the time-frequency domain, i.e.
\begin{equation}
    \label{eq:PC}
    \pC(\X) = \STFT(\iSTFT(\X)) \eqperiod
\end{equation}

In the following we present several algorithms which are based on the above projections. All of them are iterative and include projections \cref{eq:PA,eq:PC} in each iteration, however they differ in details (and in performance, as will be shown later). All these algorithms can be generalized as the iteration rule
\begin{align}
    \X^0 &= \A \exp(j\PH^0) \eqcomma \\
    \X^{m+1} &= f\left(\X^m, \pA(\X^m), \pC(\X^m)\right) \eqcomma
\end{align}
where $m$ is the iteration index. Unless a better option is available (e.g. the phase of a noisy input signal), uniform random initialization is often used, i.e. $\PH^0 \sim \mathcal{U}(-\pi,\pi)$ i.i.d. for each bin.
\subsection{Griffin-Lim algorithm (GLA)}
\vspace{-5px}
The most basic iterative algorithm has been proposed by Griffin and Lim \cite{griffinSignalEstimationModified1984} and consists of a simple iteration between the two projections:
\begin{equation}
    \label{eq:GL}
    \X^{m+1} = \pC(\pA(\X^m)) \eqperiod
\end{equation}
Similar algorithms are used in other fields based on the original algorithm proposed by Gerchberg and Saxton \cite{gerchbergPracticalAlgorithmDetermination1972}. GLA is widely used in speech and audio applications due to its simplicity. Additionally, it has a convergence guarantee (in terms of error minimization), although it is not guaranteed that it converges towards a global solution \cite{griffinSignalEstimationModified1984,sturmelSignalReconstructionSTFT2011}. A major drawback of GLA is the large number of iterations required for a reasonable reconstruction quality.
\subsection{Fast Griffin-Lim algorithm (FGLA)}
\begin{figure}
    \centering
    \includegraphics[width=.9\columnwidth]{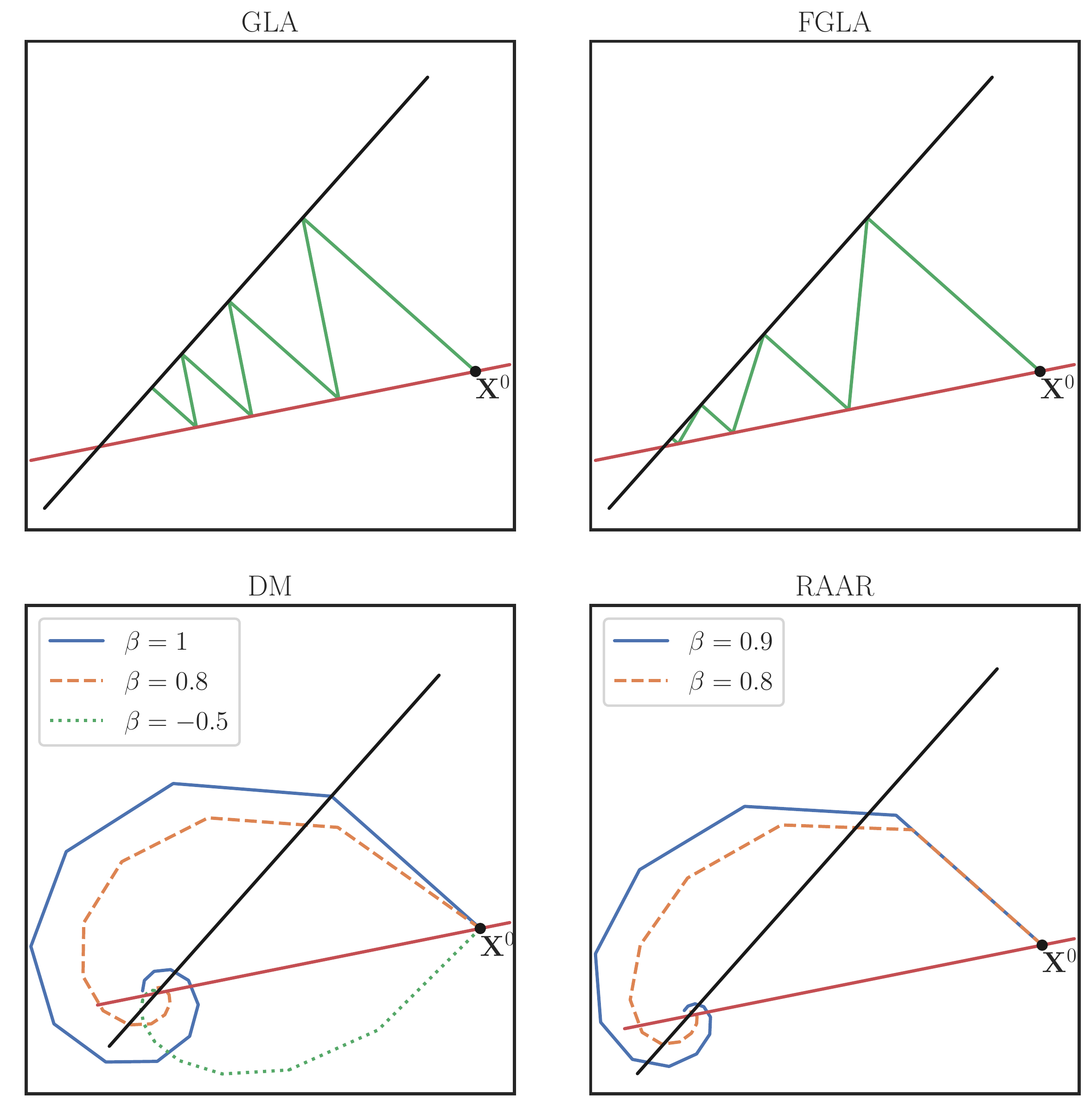}
    \caption{Illustration of different iterative algorithms on a two-dimensional convex toy problem, with $\pA$ and $\pC$ interpreted as Euclidean projections onto lines (in black and red). This figure illustrates the different trajectories taken in the simplified constraint space, but it is important to note that the actual phase retrieval problem is non-convex and non-linear. (F)GLA show a zig-zagging pattern with FGLA exhibiting a larger effective step size than GLA, whereas DM and RAAR both spiral inwards to the solution.}
    \label{fig:algos}
\end{figure}

\vspace{-5px}
In order to accelerate convergence of GLA, Perraudin et al. \cite{perraudinFastGriffinLimAlgorithm2013} proposed modifying the GLA iteration using a Nesterov acceleration scheme, yielding
\begin{align}
    \X^{m+1} &= \pC(\pA(\Y^m)) \eqcomma \label{eq:FGL1} \\ 
    \Y^{m+1} &= \X^{m+1} + \alpha_m (\X^{m+1} - \X^m) \eqcomma \label{eq:FGL2}
\end{align}
initialized with $\Y^0 = \X^0$. The parameter $\alpha_m$ has been the subject of limited analysis in \cite{perraudinFastGriffinLimAlgorithm2013}, where it has been shown that setting a constant value $\alpha_m = \alpha = 0.99$ generally leads to good results. Note that the limiting case $\alpha = 0$ is equivalent to GLA. The second term in \cref{eq:FGL2} increases the effective step size, resulting in faster convergence, as illustrated in \cref{fig:algos}. Empirical results show FGLA is superior to GLA not only in terms of convergence rate, but also in terms of reconstruction quality \cite{perraudinFastGriffinLimAlgorithm2013,masuyamaGriffinLimPhase2019}. This is also reflected in our experimental evaluation in \cref{sec:results}.
\subsection{Relaxed Averaged Alternating Reflections (RAAR)}
\vspace{-5px}
Both GLA and FGLA project the signal back and forth between the two constraint sets. This can be seen as a zigzag walk across the space spanned by these two sets. Considering the non-convexity of the phase retrieval problem, this simple strategy results in algorithms which are unable to escape sub-optimal local solutions. To mitigate this problem, one can consider walking along a more elaborate path across this space. Instead of only considering the projections onto $\setC$ and $\setA$, RAAR \cite{lukeRelaxedAveragedAlternating2004} uses the reflections about these sets to define an iteration that can escape local convergence basins. With the reflections defined as
\begin{align}
    \rA(\X) &= 2\pA(\X) - \X \eqcomma \\
    \rC(\X) &= 2\pC(\X) - \X \eqcomma
\end{align}
the RAAR iteration can be written as
\begin{equation}
    \label{eq:RAAR}
    \X^{m+1} = \frac{1}{2}\beta \left[ \X^m + \rC\left( \rA(\X^m) \right) \right] + (1-\beta) \pA(\X^m) \eqcomma
\end{equation}
with $0 < \beta \leq 1$. Recently \cite{kobayashiAcousticApplicationPhase2022}, RAAR has been found to perform well for speech signals with the best results achieved for $\beta=0.9$.
\subsection{Difference Map (DM)}
\vspace{-5px}
Another iterative algorithm which takes a more elaborate path in the constraint space is the Difference Map, proposed by Elser \cite{elserPhaseRetrievalIterated2003} as a generalization of Fienup's hybrid input-output algorithm \cite{fienupReconstructionObjectModulus1978}, commonly used in optics. The difference map shows good performance in imaging problems and has also been applied to other optimization problems \cite{elserSearchingIteratedMaps2007}. To the best of our knowledge, the Difference Map has not been previously applied to speech signals.

The basic building blocks of the algorithm remain $\pC$ and $\pA$ as in GLA, but they are used in a more elaborate fashion, resulting in a behavior which is reminiscent (but different) of RAAR (see \cref{fig:algos}). The iteration rule of DM is defined as follows:
\begin{equation}
    \label{eq:DM}
    \X^{m+1} = \X^m + \beta \left[ \pC\left(\fA(\X^m)\right) - \pA\left(\fC(\X^m)\right) \right] \eqcomma
\end{equation}
with
\begin{align}
        \fA(\X) &= \pA(\X) + \left( \pA(\X) - \X \right) / \beta \eqcomma \label{eq:DM_fA} \\
        \fC(\X) &= \pC(\X) - \left( \pC(\X) - \X \right)  / \beta  \eqperiod \label{eq:DM_fC} 
\end{align}
In contrast to RAAR, the parameter $\beta$ can take any value in $\mathbb{R} \setminus \{0\}$. However, it is often set to values close to 1, chosen experimentally as its optimal value strongly depends on signal characteristics. Note that changing the sign of $\beta$ has the effect of interchanging the projections in \cref{eq:DM,eq:DM_fA,eq:DM_fC}. For $\beta = 1$, the RAAR iteration and DM iteration are identical (this case is also equivalent to several other phase retrieval algorithms \cite{marchesiniUnifiedEvaluationIterative2007}). However, for other values of $\beta$ these algorithms are inherently different \cite{lukeRelaxedAveragedAlternating2004}. Compared to (F)GLA and RAAR, DM has a disadvantage in terms of computational complexity: For $\abs{\beta} \neq 1$, each iteration requires computing four projections instead of two.
\subsection{Alternating Direction Method of Multipliers (ADMM)}
\vspace{-5px}
Several recent contributions \cite{wenAlternatingDirectionMethods2012,liangPhaseRetrievalAlternating2018,masuyamaGriffinLimPhase2019,vialPhaseRetrievalBregman2021a} proposed treating phase retrieval directly as a constrained non-convex optimization problem and applying the alternating direction method of multipliers to it. This approach has shown good results for STFT phase retrieval as well as in other contexts. Here we consider specifically the simple unrelaxed algorithm proposed in \cite[Alg. 1]{masuyamaGriffinLimPhase2019} for STFT phase retrieval (note that similar algorithms have been previously proposed in the optics/ptychography domain, e.g. \cite{wenAlternatingDirectionMethods2012}). Using our notation, this algorithm is given by the following iteration:
\begin{align}
    \X^{m+1} &= \pA(\Z^m - \U^m) \eqcomma \label{eq:ADMM_X}\\
    \Z^{m+1} &= \pC(\X^{m+1} + \U^m) \eqcomma \label{eq:ADMM_Z}\\
    \U^{m+1} &= \U^m + \X^{m+1} - \Z^{m+1} \eqcomma \label{eq:ADMM_U} 
\end{align}
initialized with $\Z^0 = \X^0$ and $\U^0 = \mathbf{0}$. Following the derivation in \cite{yanPtychographicPhaseRetrieval2020}, we can define an auxiliary variable $\W^m = \Z^m + \U^m$, yielding
\begin{align}
    \X^{m+1} &= \pA(2\Z^m - \W^m) \eqcomma \label{eq:ADMM_X2}\\
    \Z^{m+1} &= \pC(\X^{m+1} + \W^m - \Z^m) \eqcomma \label{eq:ADMM_Z2}\\
    \W^{m+1} &= \W^m + \X^{m+1} - \Z^{m} \eqperiod \label{eq:ADMM_W} 
\end{align}
By rearranging \cref{eq:ADMM_W}, and substituting $\W^m$ into \cref{eq:ADMM_Z2}, we get
\begin{equation}
    \label{eq:ADMM_Z3}
    \Z^{m+1} = \pC(\W^{m+1}) \Rightarrow \Z^{m} = \pC(\W^{m}) \eqperiod 
\end{equation}
Finally, by combining \cref{eq:ADMM_X2,eq:ADMM_W,eq:ADMM_Z3} we obtain a single expression for the ADMM iteration:
\begin{equation}
    \label{eq:ADMM_W2}
    \W^{m+1} = \W^m + \pA(2\pC(\W^m)-\W^m) - \pC(\W^m) \eqperiod
\end{equation}
Given that $\W$ is initialized as $\W^0 = \Z^0 + \U^0 = \X^0$, the change of variables does not affect the overall algorithm. Besides providing a more compact representation, the expression in \cref{eq:ADMM_W2} allows us to compare the ADMM algorithm with other iterative algorithms. In fact, this expression is equivalent to the DM algorithm in \cref{eq:DM} with $\beta = -1$. By swapping the projections in \cref{eq:ADMM_X,eq:ADMM_Z}, it is also straightforward to define an ADMM iteration that is equivalent to DM with $\beta=1$. Note that the ADMM framework is versatile and by choosing different proximal operators, one arrives at ADMM iterations which are not equivalent to DM (e.g. the relaxed formulation in \cite{masuyamaGriffinLimPhase2019} or the divergence-based algorithms in \cite{vialPhaseRetrievalBregman2021a}).
\subsection{Proposed hybrid algorithm}
\vspace{-5px}
\label{sec:hybrid}
As shown in \cref{fig:algos}, different algorithms can take very different trajectories in the constraint space and thus also show different convergence characteristics. By first applying one algorithm, then using its result to initialize another algorithm, one can take advantage of this. Inspired by common practice in optics, we propose to start with several iterations of DM or RAAR, followed by FGLA. We expect that DM/RAAR, being able to explore a larger portion of the constraint space, will approach the vicinity of a good solution. The subsequent FGLA iterations are then likely to quickly converge into this solution. Assuming we choose DM as the first algorithm, the resulting iteration can be expressed as
\begin{equation}
    \label{eq:hybrid}
    \X^{m+1} = \begin{cases}
        \mathrm{DM}(\X^m)  & \text{if } m < M_0 \\
        \mathrm{FGLA}(\X^m)  & \text{otherwise}
    \end{cases} \eqcomma
\end{equation}
where $\mathrm{DM}(\cdot)$ and $\mathrm{FGLA}(\cdot)$ refer to the iterations defined in \cref{eq:DM} and \cref{eq:FGL1}, respectively.
This hypothesis is tested in \cref{sec:hybrid_results}, where we also study the optimal value of $M_0$
\section{Evaluation and Discussion}
\vspace{-2px}
\label{sec:results}
We perform several experiments to evaluate different aspects of the algorithms described in the previous section. All experiments are performed on clean speech data, where magnitudes are given and phase is completely discarded.
\subsection{Experimental setting}
\vspace{-5px}
For all experiments we use 100 spoken English utterances from the TIMIT corpus (gender-balanced) \cite{timit}, sampled at \SI{16}{\kilo\hertz}. The initial phase is randomly sampled from a uniform distribution as described in \cref{sec:algos}. In order to reduce the effect of random initialization on the results, we use each utterance three times with different random initialization each time and take measures to ensure the initialization is consistent between different algorithms. The STFT and iSTFT operations are performed with a frame length of \SI{32}{\ms}, a square-root Hann window and frame shift of \SI{8}{\ms}, except in \cref{sec:overlap} where the frame shift is varied. Performance of the different algorithms is measured in terms of speech quality (PESQ) and in terms of the spectral convergence (SC), defined as \cite{sturmelSignalReconstructionSTFT2011,masuyamaDeepGriffinLim2021}:
\begin{equation}
    \label{eq:sc}
    \mathrm{SC} = \frac{\left\Vert \A - \abs{\pC(\X)} \right\Vert_\mathrm{F}}{\left\Vert \A \right\Vert_\mathrm{F}}
\end{equation}
\subsection{Algorithm comparison}
\vspace{-5px}
\begin{figure*}
  \centering
  \includegraphics[width=.9\textwidth]{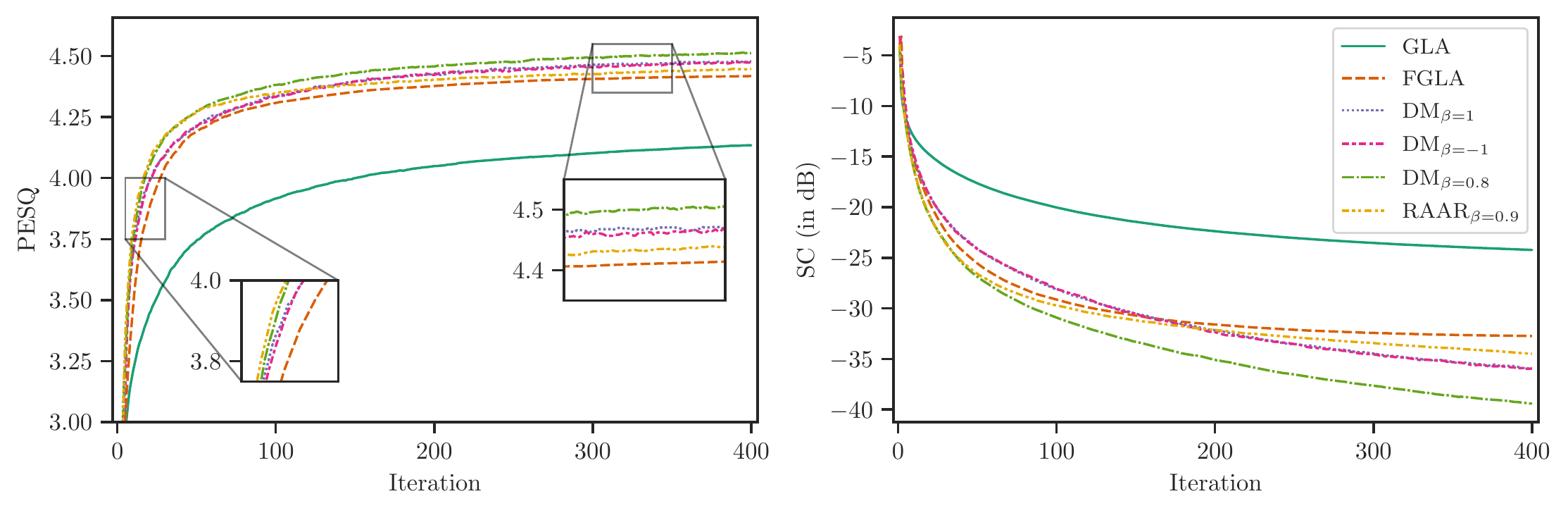}
  \caption{Mean PESQ and spectral convergence (SC) of signals reconstructed from clean magnitudes using selected algorithms from \cref{sec:algos}. Apart from the classical Griffin-Lim algorithm, all algorithms show very good reconstruction quality after 400 iterations. While both RAAR and DM perform better than FGLA, DM does exhibit consistently better perceptual results after the first few iterations especially with $\beta=0.8$. The same trend is evident in the non-perceptual SC metric.}
  \label{fig:exp1}
\end{figure*}
In \cref{fig:exp1} we compare the GLA, FGLA, RAAR and DM algorithms in terms of reconstruction quality (measured as PESQ MOS-LQO score) and spectral convergence, using the clean magnitudes and random initial phase. In order to keep the comparison simple, we only show results for selected values of $\beta$. For RAAR, $\beta=0.9$ is chosen based on results from \cite{kobayashiAcousticApplicationPhase2022}. In the case of DM, we show results for $\beta = \pm 1$ (the simplest cases which are equivalent to RAAR and the ADMM algorithm from \cite{masuyamaGriffinLimPhase2019}), as well as $\beta = 0.8$. While all algorithms reach very good reconstruction quality in 400 iterations (except GLA), we do observe certain differences. RAAR shows the steepest rise within the first few iterations and performs consistently better than FGLA (confirming results from \cite{kobayashiAcousticApplicationPhase2022}), but in later iterations it stagnates faster. The DM algorithm takes a little longer to rise but then achieves a slightly higher PESQ score than RAAR. Of all candidates, the DM algorithm with $\beta=0.8$ shows the best performance in later iterations, and also performs almost as well as RAAR in earlier iterations. The spectral convergence of each algorithm follows the same trend, implying that DM and RAAR do in fact converge against a better solution than FGLA and GLA, due to their ability to escape local sub-optimal solution. It shall be noted that while the best results are achieved with DM and $\beta =0.8$, this case requires twice as many projection operations compared to all other algorithms.
\subsection{Effect of overlap}
\vspace{-5px}
\label{sec:overlap}
\begin{figure}
    \centering
    \includegraphics[width=\columnwidth]{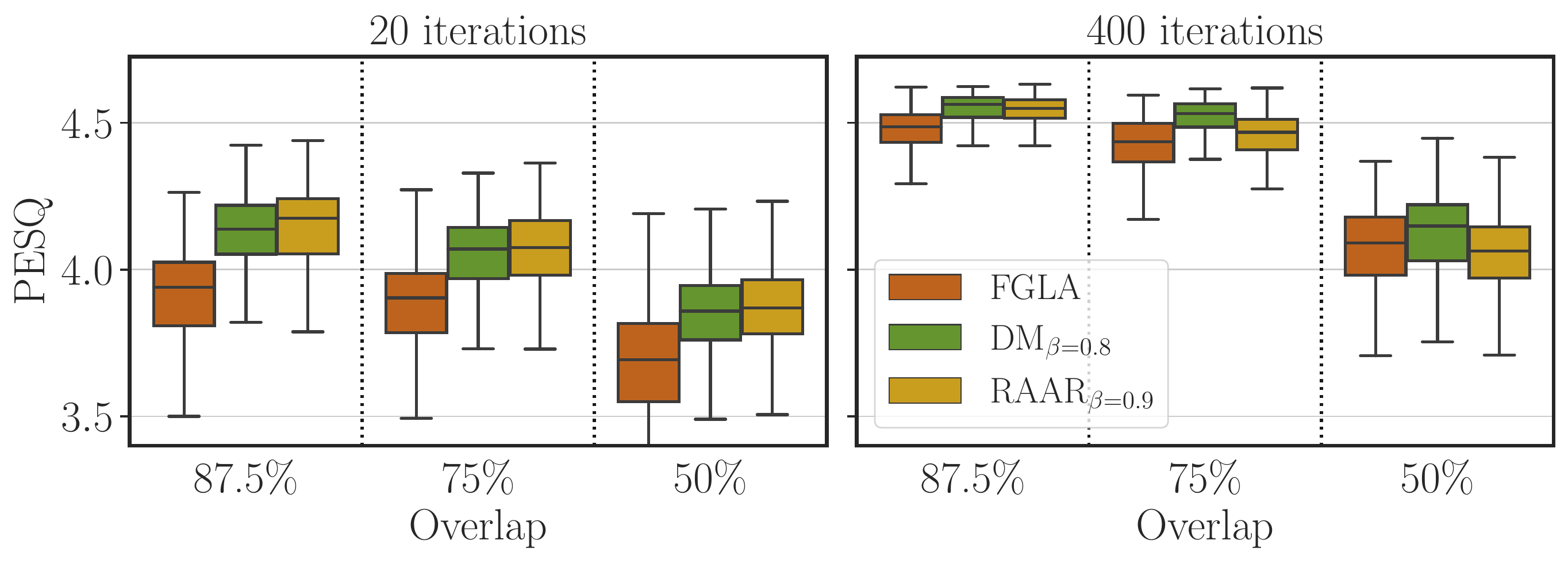}
    \caption{PESQ scores of selected algorithms after 20 and 400 iterations at different overlap ratios.}
    \label{fig:overlap}
\end{figure}
In order to gain further insight, we now examine a different performance aspect:
sensitivity to to reduced redundancy. For this purpose we repeat the same
evaluation as in the previous section, but with varying STFT frame shift,
resulting in different overlap ratios. Results are given in \cref{fig:overlap}.
Following from the definition of $\pC$, all algorithms discussed here depend
on the inherent redundancy of the STFT representation, and with reduced
redundancy, we can expect a performance drop. While this drop, as well as the
general trend of \cref{fig:exp1}, are reflected in \cref{fig:overlap}, we also
see that in later iterations, smaller overlap affects RAAR more adversely than DM.
In fact, at \SI{50}{\percent } overlap, RAAR's performance even drops slightly below
that of FGLA. We thus infer that DM and FGLA are slightly more robust to reduced
redundancy than RAAR.
\subsection{Hybrid algorithm}
\vspace{-5px}
\label{sec:hybrid_results}
\begin{figure}
    \centering
    \includegraphics[width=\columnwidth]{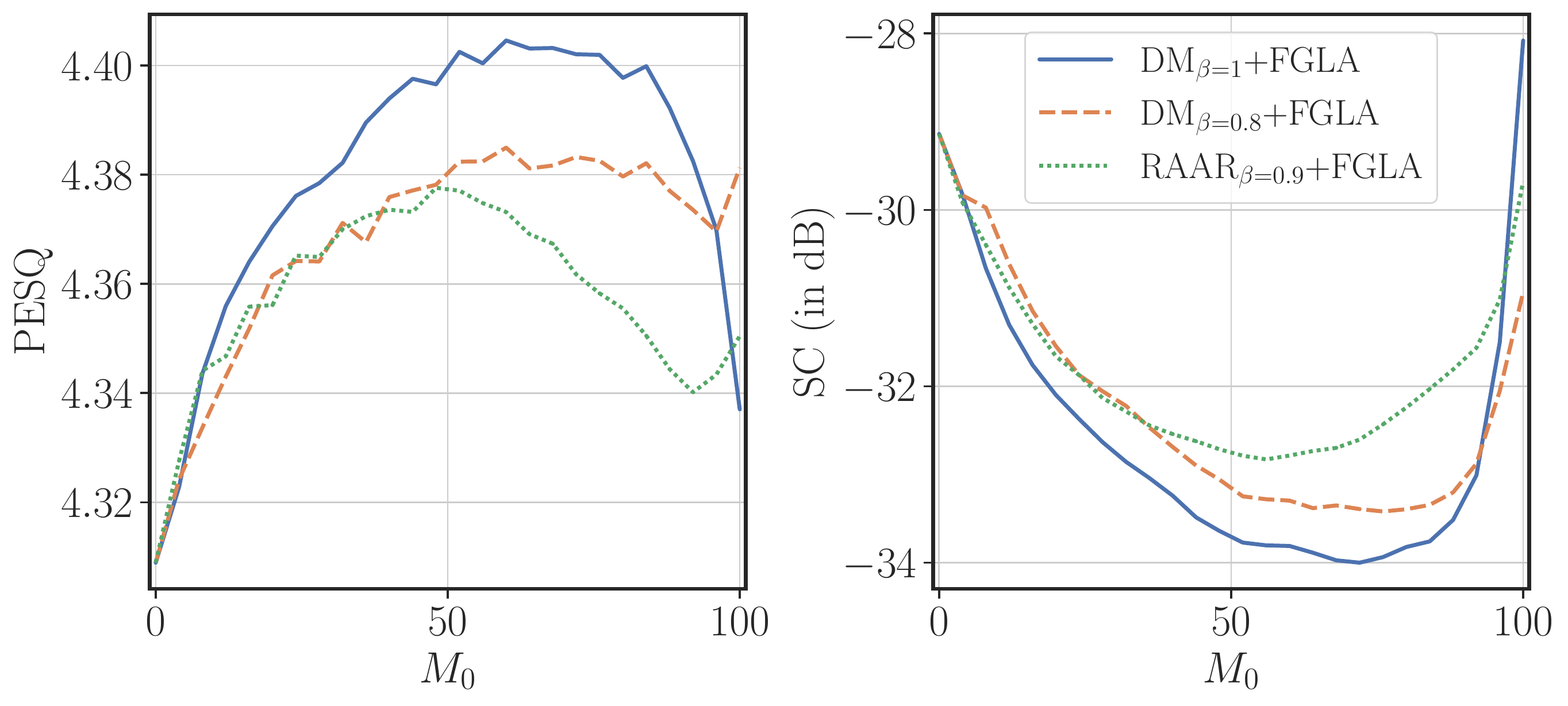}
    \caption{Mean PESQ and SC results of our proposed hybrid algorithm. Each algorithm is run for 100 iterations, of which we use DM or RAAR for $M_0$ iterations and FGLA for the rest.}
    \label{fig:hybrid_pesq_lsc}
\end{figure}
As described in \cref{sec:hybrid}, we propose a hybrid approach that combines two algorithms and benefits from the characteristics of each. To evaluate this approach, we use either RAAR or DM as the initial algorithm, followed by FGLA with a total of 100 iterations. The initial algorithm is used for $M_0$ iterations, followed by FGLA for $100-M_0$ iterations. Results of this evaluation are summarized in \cref{fig:hybrid_pesq_lsc} (note that the edge cases $M_0=0$ and $M_0=100$ correspond to pure FGLA and pure DM/RAAR, respectively). Regardless of the algorithm chosen for the initial iterations, we observe improved results using the hybrid approach. In particular, for a wide range of $M_0$, DM with $\beta=1$ even outperforms the best performing algorithm from \cref{fig:exp1} (DM with $\beta=0.8$). The best result in terms of PESQ is achieved for $M_0 = 60$. Since in this case we have $\abs{\beta} = 1$, this approach does not incur an increase in computational load.
\section{Conclusion}
\vspace{-2px}
\label{sec:conclusion}
The purpose of this paper is to show that iterative phase retrieval for speech signals is not a long-solved problem. While the Griffin-Lim algorithm and especially its accelerated variant generally perform well, recent advances from other fields in which the phase retrieval problem appears may be adapted and used for speech as well. Evaluation results indicate that given the right choice of parameters, the Difference Map algorithm can push the upper bounds of reconstruction quality. Based on the analysis of the different algorithms' convergence behavior, we propose a hybrid approach which leads to superior results without increased computational load.
\clearpage
\section{References}
\label{sec:refs}
\atColsBreak{\vskip5pt}
\printbibliography[heading=none]
\end{document}